\documentclass[12pt]{article}
\usepackage{amssymb,graphicx}
\begin{document}
%\fontsize{12}{24pt}\selectfont
\vspace*{2cm}

\begin{center}
{\Large\bf
Order-disorder model of phase transitions\\
in the DMAGaS-DMAAlS family crystals
}
\end{center}
\vspace{0.5cm}

\centerline{I.V.~Stasyuk and O.V.~Velychko}
\vspace{0.5cm}

\begin{center}
\parbox{12.5cm}{
Institute for Condensed Matter Physics\\
of the National Academy of Sciences of Ukraine,\\
1 Svientsitskii Str., 79011 Lviv, Ukraine\\
e-mail: velychko@icmp.lviv.ua
}
\end{center}
\vspace{1cm}

\begin{center}
\parbox{14cm}{
A four-state model is proposed for description of the sequence of phase
transitions in ferroelectric crystals of DMAGaS and DMAAlS type.  The
ordering processes in the subsystems of DMA groups are considered as a main
reason of such transformations. The interaction between groups in their
various orientational states is taken into account in the dipole-dipole
approximation.  Obtained thermodynamical characteristics of the model
(spontaneous polarization, occupancy of orientational states, dielectric
susceptibility, phase diagram) are in good agreement with experimental data.
The experimental fact of suppression of ferroelectric phase at increase of
hydrostatic pressure is explained under assumption that pressure
changes the difference between energies of various orientational states of
DMA groups mostly.
}
\end{center}

\section{Introduction}

A family of isomorphous crystals with ferroelectric properties to
which e.g.\ \linebreak
(CH$_3$)$_2$NH$_2$Al(SO$_4$)$_2$ $\cdot$ 6H$_2$O (DMAAlS) and
(CH$_3$)$_2$NH$_2$Ga(SO$_4$)$_2$ $\cdot$ 6H$_2$O (DMAGaS) belong to is
experimentally studied during the past decade. A peculiar feature of
the family is a possible existence of the crystal in three different
phases at change of temperature: at room temperature the crystal is
paraelectric, at lowering of temperature it sequentially becomes
ferroelectric and antiferroelectric
(Pietraszko, \L{}ukaszewicz and Kirpichnikova, 1993;
Pietraszko and \L{}ukaszewicz, 1994;
Pietraszko, \L{}ukaszewicz and Kirpichnikova, 1995;
Tchukvinskyi, Cach and Czapla, 1998).
For example, it has been measured $T_{c1}=136$~K and $T_{c2}=113$~K
by Tchukvinskyi {\it et al.} (1998)
(or 122~K and 114~K correspondingly
(Pietraszko {\it et al.}, 1995)) for
DMAGaS crystal, but it has been only found $T_{c1}=150$~K
(Pietraszko {\it et al.}, 1993; Pietraszko {\it et al.}, 1994; Pietraszko {\it et al.}, 1995)
for DMAAlS. The phase transition between the
ferroelectric and antiferroelectric phases is of the first order
(Pietraszko {\it et al.}, 1995; Tchukvinskyi {\it et al.}, 1998).
There are evidences, supplied by
optical, ultrasonic, pyroelectric, dilatometric and dielectric measurements,
that the phase transition paraelectric -- ferroelectric is of the first
order close to the second one in DMAGaS and of the second order in DMAAlS.
Crystallographic analysis shows that in all three phases the crystal
belongs to monoclinic space groups: the high-temperature paraelectric phase
has P2$_1$/n space group
(Pietraszko {\it et al.}, 1993; Pietraszko {\it et al.}, 1994), ferroelectric and
antiferroelectric ones have Pn
(Pietraszko {\it et al.}, 1994; Pietraszko {\it et al.}, 1995) and P2$_1$
(Pietraszko {\it et al.}, 1995) groups respectively. It should be mentioned that
low-symmetry space groups are subgroups of the high-symmetry group
obtained by loss of rotation axis and mirror plane respectively
(point symmetry group 2/m changes to m or 2).

\begin{figure}
\centerline{
\includegraphics[width=0.4\textwidth]{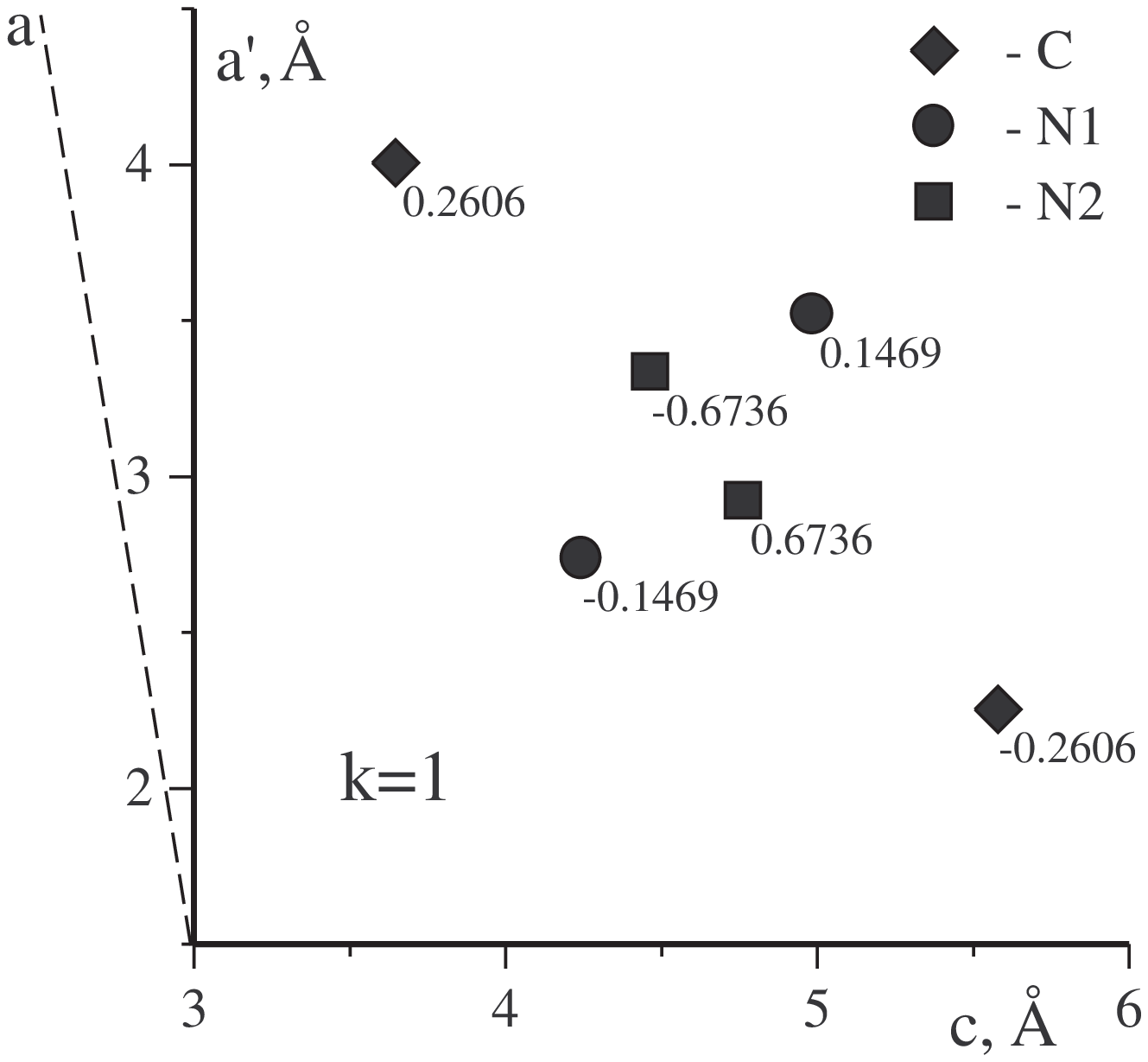}
\hspace{2em}
\includegraphics[width=0.4\textwidth]{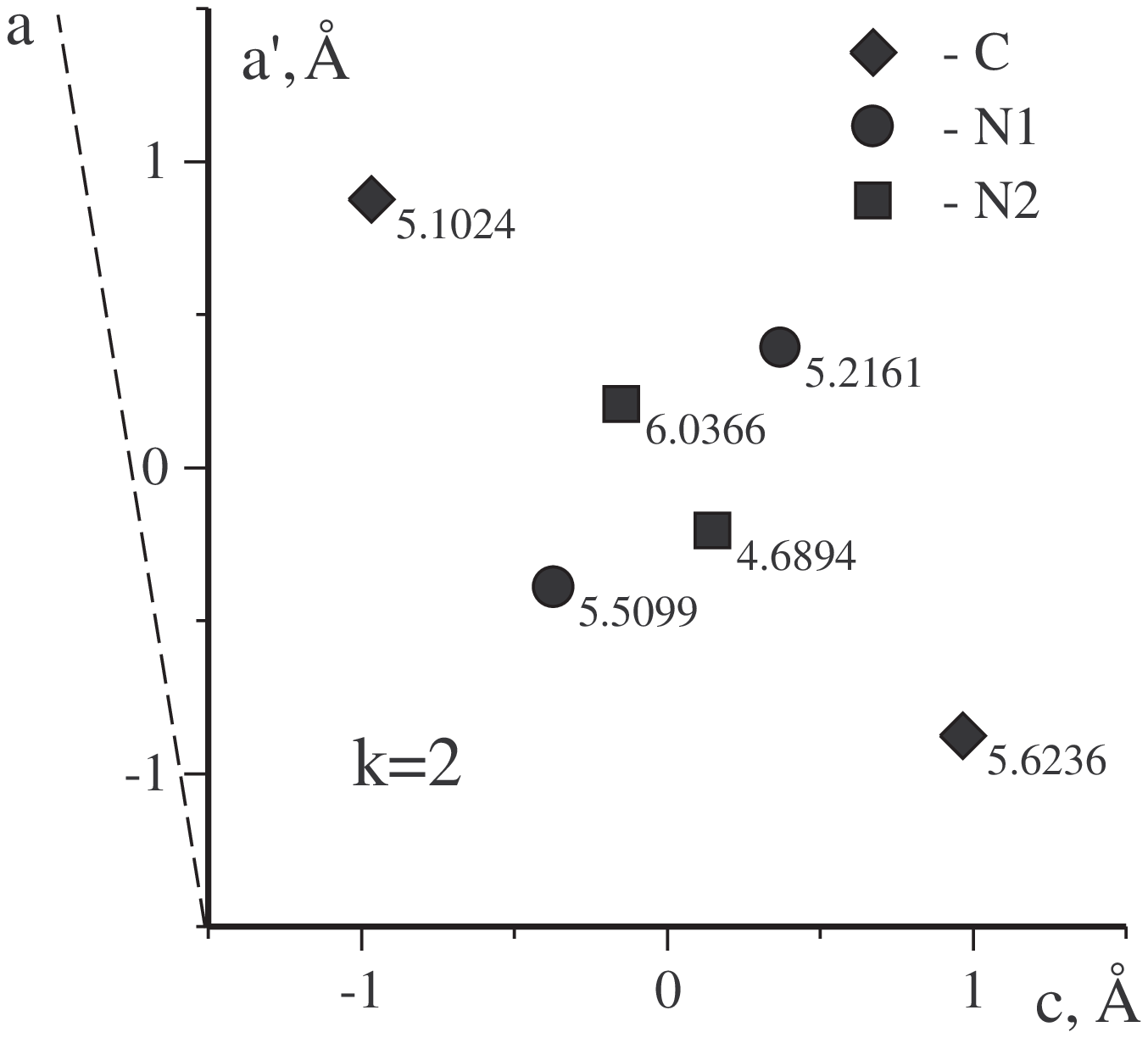}
}
\caption{Projection of N and C atoms in the dimethylammonium groups of
DMAGaS crystal at ambient temperature
(Pietraszko {\it et al.}, 1993)
onto the $ac$
plane (the $b$ coordinate is indicated for each atom; $k$ is the index of DMA
sublattices).}
\label{xzprj}
\end{figure}
%\vspace{3ex}

\begin{figure}
\centerline{\includegraphics[width=0.45\textwidth]{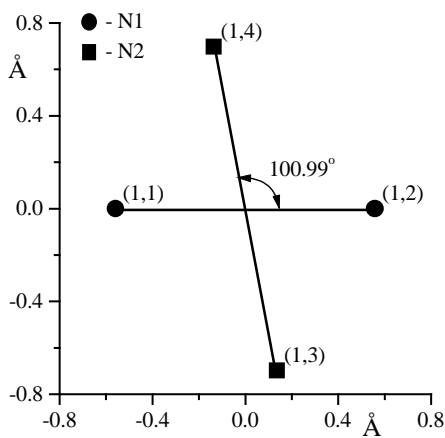}}
\caption{
Projection of N atoms onto the plane
perpendicular to the C--C$'$ axis of the dimethylammonium group
for the same crystal as in Figure~\protect\ref{xzprj}.
}
\label{cross}
\end{figure}

Certain ferroelectric compounds have structural elements which
reorientation leads to polarisation of the crystal. Changes in the rest of the
crystal are insignificant and can be neglected in description of the
phenomenon. Crystals DMAAlS and DMAGaS belong to this type of
ferroelectrics. Here the element, which can be reoriented, is the
dimethylammonium cation (more strictly NH$_2$ group)
(Pietraszko {\it et al.}, 1995;
Sobiestinskas, Grigas, Andreev and Varikash, 1992;
Kazimirov {\it et al.}, 1998;
Dolin\v{s}ek {\it et al.}, 1999).
This element can
occupy four equilibrium positions related in pairs by inversion
centre forming a slightly deformated cross (Figures~\ref{xzprj} and~\ref{cross}).
In the paraelectric phase a site in one pair (($k,1$) and ($k,2$);
$k=1,2$) is occupied with
probability 40\% and in another (($k,3$) and ($k,4$))
with probability 10\% at 300~K (Pietraszko {\it et al.}, 1993).
Instead of the presented in Figure~\ref{xzprj} oblique-angled
crystallographic coordinates we shall use rectangular ones hereinafter. Their OY~axes
coincides with the $b$~axis, the OX~axis is directed along the ferroelectric
axis (N1--N1) and the OZ~axis is chosen to make planes XZ and $ac$
identical.

According to the experimental data (Pietraszko {\it et al.}, 1995) in the
ferroelectric and antiferroelectric phases only positions in the first pair
are occupied with appropriate ordering.

In the ferroelectric phase positions (1,2) and (2,2) (or (1,1) and (2,1))
are mainly occupied while in the antiferroelectric state occupancies of
positions (1,2) and (2,1) (or (1,1) and (2,2) respectively) prevail
(Pietraszko {\it et al.}, 1993). In
(Stasyuk and Velychko, 2000a)
we have performed the
symmetry analysis of functions being linear combinations of position
occupancies of NH$_{2}$ groups in an elementary cell. Symmetrized linear
combinations transforming according to irreducible representations of
point symmetry group 2/m of the high-temperature phase are presented in
Table~\ref{Averg1}.  Combinations $y_{+}$, $u_{-}$ ($y_{-}$, $u_{+}$)
belonging to representation $B_u$ ($A_u$) correspond to orderings appearing
in the ferroelectric (antiferroelectric) phase and form order parameters
for these phases. Existing orderings are of a mixed nature: ferroelectric
ordering in positions ($k,1$) and ($k,2$) (along the OX axis) is accompanied
by antiferroelectric one in ($k$,3) and ($k$,4) (along the OY axis) and vice
versa.

\begin{table}%[h]
\caption{Conversion table of occupancies of orientational states into
symmetrized ones which correspond to irreducible representations of the
point symmetry group 2/m.}
\label{Averg1}
\begin{center}
\renewcommand{\arraystretch}{0}
\begin{tabular}{|c|c||c|c|c|c|c|c|c|c|}
\hline
&&(1,1)&(1,2)&(1,3)&(1,4)&(2,1)&(2,2)&(2,3)&(2,4)\strut\\
\hline
\rule{0pt}{2pt}&&&&&&&&&\\
\hline
\raisebox{-1.7ex}[0pt][0pt]{A$_g$}
      & $x_+$& 1/2&  1/2&  0&  0&  1/2&  1/2&  0&  0\strut\\
\cline{2-10}
      & $z_+$& 0&  0&  1/2&  1/2&  0&  0&  1/2&  1/2\strut\\
\hline
\raisebox{-1.7ex}[0pt][0pt]{B$_g$}
      & $x_-$& 1/2&  1/2&  0&  0& -1/2& -1/2&  0&  0\strut\\
\cline{2-10}
      & $z_-$& 0&  0&  1/2&  1/2&  0&  0& -1/2& -1/2\strut\\
\hline
\raisebox{-1.7ex}[0pt][0pt]{B$_u$}
      & $y_+$& 1/2& -1/2&  0&  0&  1/2& -1/2&  0&  0\strut\\
\cline{2-10}
      & $u_-$& 0&  0&  1/2& -1/2&  0&  0& -1/2&  1/2\strut\\
\hline
\raisebox{-1.7ex}[0pt][0pt]{A$_u$}
      & $y_-$& 1/2& -1/2&  0&  0& -1/2&  1/2&  0&  0\strut\\
\cline{2-10}
      & $u_+$& 0&  0&  1/2& -1/2&  0&  0&  1/2& -1/2\strut\\
\hline
\end{tabular}
\renewcommand{\arraystretch}{1}
\end{center}
\end{table}

Proceeding from these facts we have proposed microscopic four-state model
(Stasyuk {\it et al.}, 2000a)
for description of thermodynamics of order-disorder phase transitions in the
subsystem of reorientationable DMA groups. The phase transition between the
paraelectric and ferroelectric phases in the considered crystals was
described and criteria of order of this transition were established in the
approach
of a direct interaction between variously
oriented groups (a site-site approach).
The results obtained can not be
extended on a wide temperature range including e.g. the region of the
low-temperature phase transition in DMAGaS because total occupancies of
``longitudinal'' positions 1 and 2
(and ``transverse'' positions 3 and 4
respectively) have been assumed as temperature independent parameters of
theory.
Nevertheless in this work order parameters corresponding
to anti- and ferroelectric orderings have been constructed what allowed to
describe the phase transition between ferro- and paraelectric phases and to
establish criteria of its order.

Here a continuation of this work is presented where an
approach of dipole-dipole interaction between ionic groups is used;
temperature change of total occupancies is taken into
account as well as the effect of reorientational hopping of groups between
positions ($k,p$). Ordering of ionic groups corresponding to order
parameters of both $B_u$ and $A_u$ symmetry is described in the mean field
approximation.  Equilibrium states are investigated, temperature dependences
of order parameters (spontaneous polarization) and occupancies of positions
of NH$_{2}$ groups are studied. Both phase transitions (at temperatures
$T_{c1}$ and $T_{c2}$) are described; the regions of existence of
para-, ferro- and antiferroelectric phases depending on temperature and model
parameters are established. Observed by experiment changes of picture of
phase transitions in DMAGaS crystal under the influence of hydrostatic
pressure are interpreted in the framework of microscopic  description.

\section{Hamiltonian of the model}

Let us formulate Hamiltonian of the subsystem of DMA groups on the basis of
four orientational states with taking into account of differences of energies of
nonequivalent configurations, possibility of orientational hopping and interaction
between groups in dipole approximation.
\begin{equation}
H = -\sum_{nk} \sum_{\alpha} E_{\alpha} D_{nk}^{\alpha} +
\sum_{nk} \sum_{ss'} \lambda^{ss'}_{k} X_{nk}^{ss'}
- \frac{1}{2} \sum_{nn'}\sum_{kk'}
\sum_{\alpha\beta} \Psi_{\alpha\beta}^{kk'} (nn') D_{nk}^{\alpha}
D_{n'k'}^{\alpha'} \: ,
\label{eq1}
\end{equation}
where
\begin{equation}
\hat{\lambda}_{1} = \left(\begin{array}{cccc}
\varepsilon_{1} & 0 & \Omega_{1} & \Omega_{2} \\
0 & \varepsilon_{1} & \Omega_{2} & \Omega_{1} \\
\Omega_{1} & \Omega_{2} & \varepsilon_{2} & 0 \\
\Omega_{2} & \Omega_{1} & 0 & \varepsilon_{2} \\
\end{array} \right), \quad
\hat{\lambda}_{2} = \left(\begin{array}{cccc}
\varepsilon_{1} & 0 & \Omega_{2} & \Omega_{1} \\
0 & \varepsilon_{1} & \Omega_{1} & \Omega_{2} \\
\Omega_{2} & \Omega_{1} & \varepsilon_{2} & 0 \\
\Omega_{1} & \Omega_{2} & 0 & \varepsilon_{2}\\
\end{array}\right),
\label{eq2}
\end{equation}
\begin{equation}
D_{nk}^{x} = d_x (X_{nk}^{22} - X_{nk}^{11}), \quad
D_{nk}^{y} = d_y (X_{nk}^{44} - X_{nk}^{33}), \quad
D_{nk}^{z} =0,
\label{eq3}
\end{equation}
$X_{nk}^{ss'}$ is the Hubbard operator describing transition from the state $s'$ on
the state $s$ of the DMA complex in the lattice site $n$ and the sublattice $k$
($X_{nk}^{ss'}=|nk,s\rangle \langle nk,s'|$), $D_{nk}^{\alpha}$ is the
$\alpha$-component of the dipole moment of complex,
$E_{\alpha}$ is the corresponding component of external electric field,
$\Psi_{\alpha\beta}^{kk'}(nn')$ is the energy of the
dipole interaction, $\varepsilon_{1}$ and $\varepsilon_{2}$ are the energies
of DMA groups in positions ($k,1$), ($k,2$) and ($k,3$),
($k,4$) respectively, $\Omega_{1}$ and $\Omega_{2}$ are the parameters of
the reorientational hopping.

Here the coordinate system with the X axis along the axis of
spontaneous polarization of the crystal and the Y axis perpendicular to
the DMA plane (parallel to the crystallographic $b$ axis) is used.

It should be mentioned that the Hamiltonian (\ref{eq1}) simplifies the
real structure of DMA complexes in DMAGaS crystals, which are oriented not
exactly along the chosen axes (as assumed in definitions of $D_{nk}^{\alpha}$
(\ref{eq3})). Next, in our case the dipole-dipole interaction is only the main
term of the multipole expansion.
Nevertheless, as will be shown below, the model described by the Hamiltonian
(\ref{eq1}) can give a satisfactory explanation of the behaviour
of DMAGaS and DMAAlS-type crystals despite of simplifications made.

The Hamiltonian (\ref{eq1}) can be rewritten to separate the mean field part
\begin{eqnarray}
H&=&NH_{\rm C} + H_{\rm MF} + H',
\label{eq4}
\\
H_{\rm C}&=& \frac{1}{2} \sum_{kk'} \sum_{\alpha\beta}
             \psi_{\alpha\beta}^{kk'}
             \langle D_{k}^{\alpha}\rangle \langle D_{k'}^{\beta} \rangle,
\label{eq5}
\\
H_{\rm MF} &=& \sum_{nk}\sum_{ss'} \lambda^{ss'}_{k} X_{nk}^{ss'}  -
    \sum_{nk} \sum_{\alpha} (F_{\alpha}^{k}+E_{\alpha}) D_{nk}^{\alpha},
\label{eq6}
\end{eqnarray}
where
$$
F_{\alpha}^{k} = \sum_{k'}\sum_{\beta} \psi_{\alpha\beta}^{kk'}
                     \langle D_{k'}^{\beta}\rangle\:, \quad
\psi_{\alpha\beta}^{kk'} = \sum_{n'} \Psi_{\alpha\beta}^{kk'} (nn');
$$
$N$ is the total number of sites. Further considerations will be
restricted to the mean field approximation (MFA) and hence the fluctuative
term $H'$ is neglected.

It is convenient to make a unitary
transformation to linear combinations of dipole moments
\begin{equation}
\langle D_{\eta}\rangle = \sum_{k\alpha} U_{\eta,k\alpha} \langle
D_{k}^{\alpha}\rangle
\label{eq8.1}
\end{equation}
transforming according to the irreducible
representations of the
high-temperature symmetry group of the crystal (cf. Appendix A).
After that   the MFA Hamiltonian looks like
\begin{eqnarray}
H_{\rm MFA} &=& NH_{\rm C} + H_{\rm MF}\:,
\label{eq8a}
\\
H_{\rm C} &=& \frac{1}{2} \sum_{\eta\eta'}
              \langle \tilde{D}_{\eta}\rangle \psi_{\eta\eta'}
              \langle\tilde{D}_{\eta'}\rangle,
\label{eq8}
\\
H_{\rm MF} &=& \sum_{nk} \sum_{ss'} B_{k}^{ss'} X_{nk}^{ss'},
\label{eq9}
\end{eqnarray}
where
\begin{eqnarray}
&&
B_{k}^{ss'}= \lambda^{ss'}_{k} -
               \delta_{ss'} \sum_{\eta}
                \left(
                   \sum_{\alpha} U_{\eta,k\alpha} d_{k\alpha}^{ss}
                \right)
                (\tilde{F}_{\eta}+\tilde{E}_{\eta}),
\label{eq10}
\nonumber\\
&&
\tilde{F}_{\eta} = \sum_{\eta'} \psi_{\eta\eta'}
                     \langle\tilde{D}_{\eta'}\rangle,
\quad
\psi_{\eta\eta'} = \sum_{kk'} \sum_{\alpha\alpha'}
U_{\eta,k\alpha} \psi_{kk'}^{\alpha\alpha'} U_{\eta',k'\alpha'}\:;
\nonumber\\
&&
\tilde{E}_{\eta} = \sum_{k\alpha} U_{\eta,k\alpha} E_{\alpha}\:, \quad
\tilde{E}_1 = \tilde{E}_3 = \tilde{E}_5 = 0, \;
\nonumber\\
&&
\tilde{E}_2 = \sqrt{2} E_y \,,\;
\tilde{E}_4 = \sqrt{2} E_x \,,\; \tilde{E}_6 = \sqrt{2} E_z \,.
\label{eq11}
\end{eqnarray}

Symmetrized averages of dipole moments form two pairs
belonging to different irreducible representations:
\begin{eqnarray}
& A_{u}:\qquad
\langle\tilde{D}_{1}\rangle =
(\langle D_{1}^{x}\rangle - \langle D_{2}^{x}\rangle)/\sqrt{2}, \quad
\langle\tilde{D}_{2}\rangle =
(\langle D_{1}^{y}\rangle + \langle D_{2}^{y}\rangle)/\sqrt{2};
&
\label{eq12}
\\
& B_{u}:\qquad
\langle\tilde{D}_{4}\rangle =
(\langle D_{1}^{x}\rangle + \langle D_{2}^{x}\rangle)/\sqrt{2}, \quad
\langle\tilde{D}_{5}\rangle =
(\langle D_{1}^{y}\rangle - \langle D_{2}^{y}\rangle)/\sqrt{2}.
&
\label{eq13}
\end{eqnarray}
Matrix $\psi_{\eta\eta'}$ has a block-diagonal structure with nonzero blocks
belonging to the given representation (see Appendix A).

As follows from the expressions (\ref{eq12}) and (\ref{eq13}) the
antiferroelectric ordering along the X axis is accompanied by the
ferroelectric one along the Y axis and vice versa.

In general case
$\Omega_{i}\neq 0$ the diagonalization procedure should be applied
to the matrix $\hat{B}_{k}$ in the expression (\ref{eq9}):
\begin{equation}
H_{\rm MF} = \sum_{r}\tilde{\lambda}_{k}^{r} \tilde{X}_{nk}^{rr},
\label{eq14}
\end{equation}
where
\begin{equation}
\tilde{\lambda}_{k}^{r} = \sum_{ss'} V_{sr}^{k} B_{k}^{ss'} V_{s'r}^{k},
\quad \tilde{X}_{nk}^{rr} = \sum_{ss'} V_{sr}^{k} X_{nk}^{ss'} V_{s'r}^{k}
\label{eq15}
\end{equation}
and
\begin{equation}
\langle D_{k}^{\alpha}\rangle = \sum_{r}  \tilde{d}_{k\alpha}^{rr}
\langle\tilde{X}_{k}^{rr}\rangle, \quad \tilde{d}_{k\alpha}^{rr} =
\sum_{ss'} V_{sr}^{k} d_{\alpha}^{ss'} V_{s'r}^{k}.
\label{eq16}
\end{equation}
Now all necessary thermodynamical functions could be derived from the
Hamiltonian (\ref{eq8a}) with corresponding terms expressed by (\ref{eq8})
and (\ref{eq14}).

\section{Thermodynamics of the model}

The partition function of the model in general form is presented as
\begin{equation}
Z=Z_{1}^{N}Z_{2}^{N} \exp (-\beta NH_{\rm C}), \qquad \beta=1/k_{\rm B}T,
\label{eq17}
\end{equation}
where
\begin{eqnarray}
Z_{k} &=& \sum_{s}\exp(-\beta\tilde{\lambda}_{k}^{s}), \quad k=1,2;
\label{eq18}
\\
H_{\rm C} &=&
\frac{1}{2} \left(a_{1}\langle\tilde{D}_{1}\rangle^{2}
+d_{1}\langle\tilde{D}_{2}\rangle^{2}
+2b_{1} \langle\tilde{D}_{1}\rangle\langle\tilde{D}_{2}\rangle \right.
\nonumber
\\
&&\left.
{}\quad +a_{2}\langle\tilde{D}_{4}\rangle^{2}
+d_{2}\langle\tilde{D}_{5}\rangle^{2}
+2b_{2}\langle\tilde{D}_{4}\rangle\langle\tilde{D}_{5}\rangle\right).
\label{eq19}
\end{eqnarray}
As a result free energy per particle looks like
\begin{equation}
F=-\frac{1}{\beta N} \ln Z = H_{\rm C} - \frac{1}{\beta} \ln Z_{1} -
\frac{1}{\beta} \ln Z_{2}\,.
\label{eq20}
\end{equation}
Now one can calculate averages
\begin{equation}
\langle X_{k}^{rr}\rangle = \exp (-\beta\tilde{\lambda}_{k}^{r})
          \Bigm/ \sum_{s}\exp(-\beta\tilde{\lambda}_{k}^{s})
\label{eq21}
\end{equation}
and after substitution into formulae  (\ref{eq12}), (\ref{eq13}) and
(\ref{eq16}) selfconsistency equations for $\langle\tilde{D}_{n}\rangle$
can be obtained.

In the case $\Omega_{1}=\Omega_{2}=0$ the matrix $\hat{B}_{k}$
is diagonal and expressions for thermodynamical functions are more simple.
For example, the set of selfconsistency equations looks like
\begin{eqnarray}
\nonumber
\langle\tilde{D}_{1,4}\rangle &=&
\frac{d_x}{\sqrt{2}} \exp(\beta\Delta)
\Bigl[
{\cal Z}_{1}^{-1} \sinh(\beta\varkappa_{1}^{x}) \mp
{\cal Z}_{2}^{-1} \sinh(\beta\varkappa_{2}^{x})
\Bigr],
\\
\langle\tilde{D}_{2,5}\rangle &=&
\frac{d_y}{\sqrt{2}} \exp(-\beta\Delta)
\Bigl[
{\cal Z}_{1}^{-1} \sinh(\beta\varkappa_{1}^{y}) \pm
{\cal Z}_{2}^{-1} \sinh(\beta\varkappa_{2}^{y})
\Bigr],
\label{eq22}
\end{eqnarray}
where
\begin{eqnarray}
\nonumber
{\cal Z}_{k} &=&
  \exp(\beta\Delta) \cosh(\beta\varkappa_{k}^{x}) +
  \exp(-\beta\Delta) \cosh(\beta\varkappa_{k}^{y}), \quad k=1,2,
\\
\nonumber
\varkappa_{k}^{x} &=& \frac{d_x}{\sqrt{2}}
\Bigl[
(-1)^{\delta_{k,2}}
(a_{1}\langle\tilde{D}_{1}\rangle + b_{1} \langle\tilde{D}_{2}\rangle) +
a_{2} \langle\tilde{D}_{4}\rangle + b_{2} \langle\tilde{D}_{5}\rangle  +
\sqrt{2} E_x
\Bigr],
\\
\varkappa_{k}^{y} &=& \frac{d_y}{\sqrt{2}}
\Bigl[
b_{1}\langle\tilde{D}_{1}\rangle + d_{1} \langle\tilde{D}_{2}\rangle +
(-1)^{\delta_{k,2}}
(b_{2} \langle\tilde{D}_{4}\rangle + d_{2} \langle\tilde{D}_{5}\rangle) +
\sqrt{2} E_y
\Bigr],
\nonumber
\\
\nonumber
&&
\varepsilon=(\varepsilon_{2}+\varepsilon_{1})/2, \quad
\Delta = (\varepsilon_{2}-\varepsilon_{1})/2.
\end{eqnarray}

Components of static dielectric susceptibility are determined in usual way
\begin{eqnarray}
\chi_{x} &=& \frac{\partial P_x}{\partial E_x} =
\frac{1}{v_c}
\frac{\partial \langle D_1^x + D_2^x \rangle}{\partial E_x} =
\frac{\sqrt{2}}{v_c} \,
\frac{\partial \langle \tilde{D}_4^x \rangle}{\partial E_x},
\nonumber\\
\chi_{y} &=&  \frac{\partial P_y}{\partial E_y} =
\frac{1}{v_c}
\frac{\partial \langle D_1^y + D_2^y \rangle}{\partial E_y} =
\frac{\sqrt{2}}{v_c} \,
\frac{\partial \langle \tilde{D}_2^x \rangle}{\partial E_y},
\label{scdef}
\end{eqnarray}
where $P_x$ and $P_y$ are the components of spontaneous polarization and $v_c$
is the volume of an elementary cell.
After straightforward but cumbersome calculations one can derive them from
set of equations (\ref{eq22}) in the explicit form
\begin{eqnarray}
\chi_{x} &=&
\frac{1}{v_c}
\frac{2[\Theta K_x - d_2 (K_x K_y - L^2)]}%
{\Theta^2 + \Theta(2 b_2 L - a_2 K_x -d_2 K_y) +
(a_2 d_2 - b_2^2)(K_x K_y - L^2)},
\nonumber\\
\chi_{y} &=&
\frac{1}{v_c}
\frac{2[\Theta K_y - a_1 (K_x K_y - L^2)]}%
{\Theta^2 + \Theta(2 b_1 L - a_1 K_x -d_1 K_y) +
(a_1 d_1 - b_1^2)(K_x K_y - L^2)},
\label{susc}
\end{eqnarray}
where
\begin{eqnarray}
&&
K_{\alpha} = d_{\alpha}^2 N_{\alpha} - \langle D_1^{\alpha} \rangle^2,
\quad
\alpha = x,y;\qquad L= \langle D_1^x \rangle \langle D_1^y \rangle;
\nonumber\\
&&
N_x = \langle X^{11}+X^{22} \rangle, \quad
N_y = \langle X^{33}+X^{44} \rangle, \quad \Theta=k_{\rm B}T.
\end{eqnarray}

\section{Numerical results}

The model under consideration demonstrates a complicated thermodynamical
behaviour and a some numbers of qualitatively different phase sequences
dependently on parameter values. Further study will be devoted to the
case of the sequence of phase transitions which is characteristic to the DMAGaS
crystal (antiferroelectric $\rightarrow$ ferroelectric $\rightarrow$
paraelectric phases with temperature increase).

Such a sequence of phase transition is achieved at the next relationships
between parameters
\begin{eqnarray}
&&
0<b_{1},d_{1},a_{1}<b_{2},d_{2}; \nonumber\\
&&
a_{1}>a_{2}>0.
\end{eqnarray}
If the above nonequalities are fulfilled, antiparallel orientation of
$D_{x}$- and $D_{y}$-dipoles of neighbour  DMA groups is energetically
favourable. At low temperatures (when occupancies of positions ($k,3$) and
($k,4$) tend to zero) the antiferroelectric phase related with antiparallel
dipoles $D_{x}$ is realized. In the result of increased probability  of
appearance of groups in the mentioned positions at temperature growing the
interaction with participation of $D_{y}$ dipoles becomes significant. The
phase transition to the phase with their antiferroelectric orientation occurs
when respective interactions become prevailing. This phase is
simultaneously ferroelectric along the OX axis.

There are two other more subtle conditions: for a correct description of the
DMAGaS crystal the high-temperature phase transition should be of the first
order and the ratio $(\Theta_{c1}-\Theta_{c2})/\Theta_{c1}$ should correspond
to the experimental one.

\begin{figure}
\begin{center}
\setlength{\unitlength}{0.1\textwidth}
\begin{picture}(10,8)
\put(0,4){\includegraphics[width=0.5\textwidth]{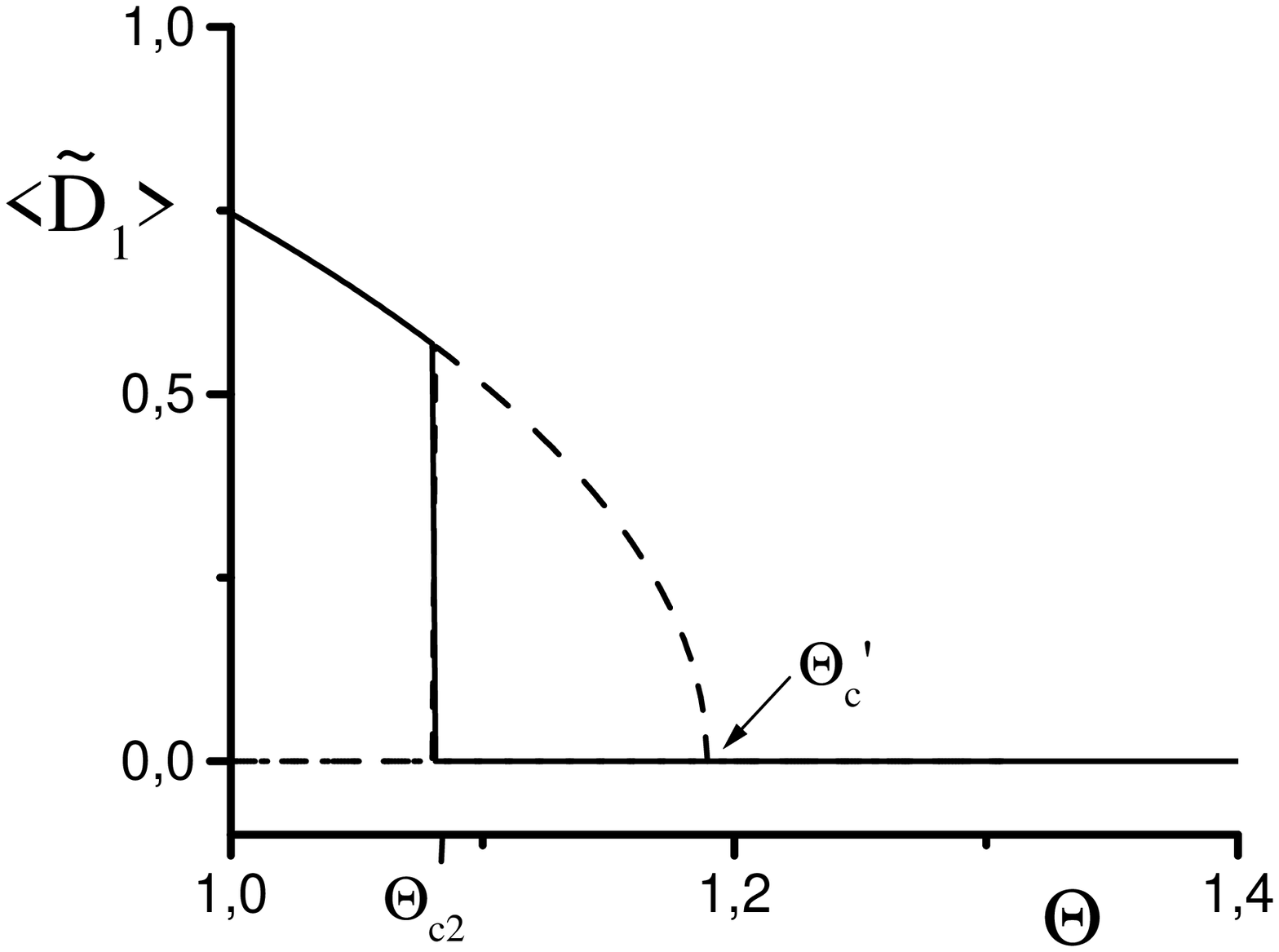}}
\put(5,4){\includegraphics[width=0.5\textwidth]{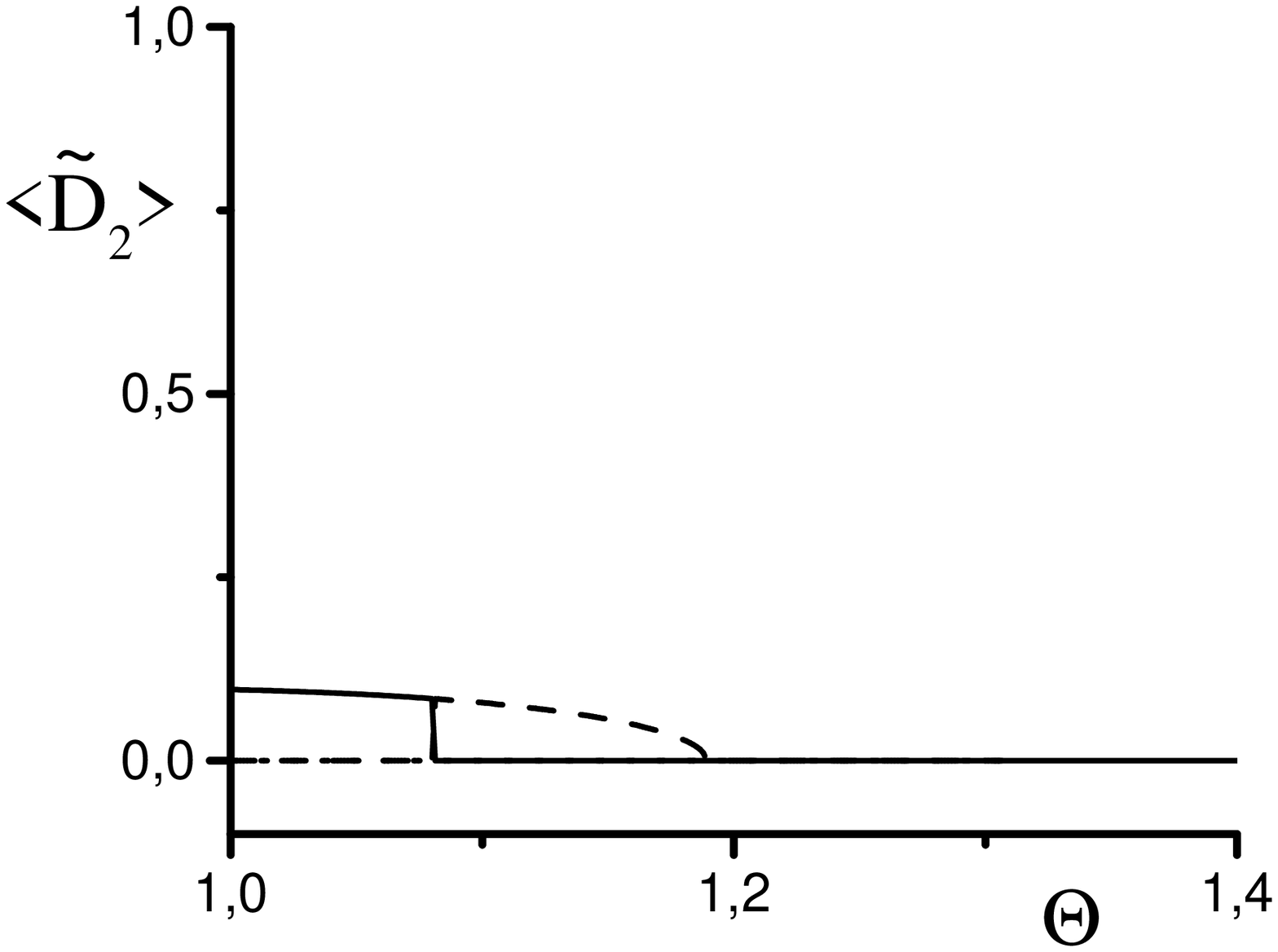}}
\put(0,0){\includegraphics[width=0.5\textwidth]{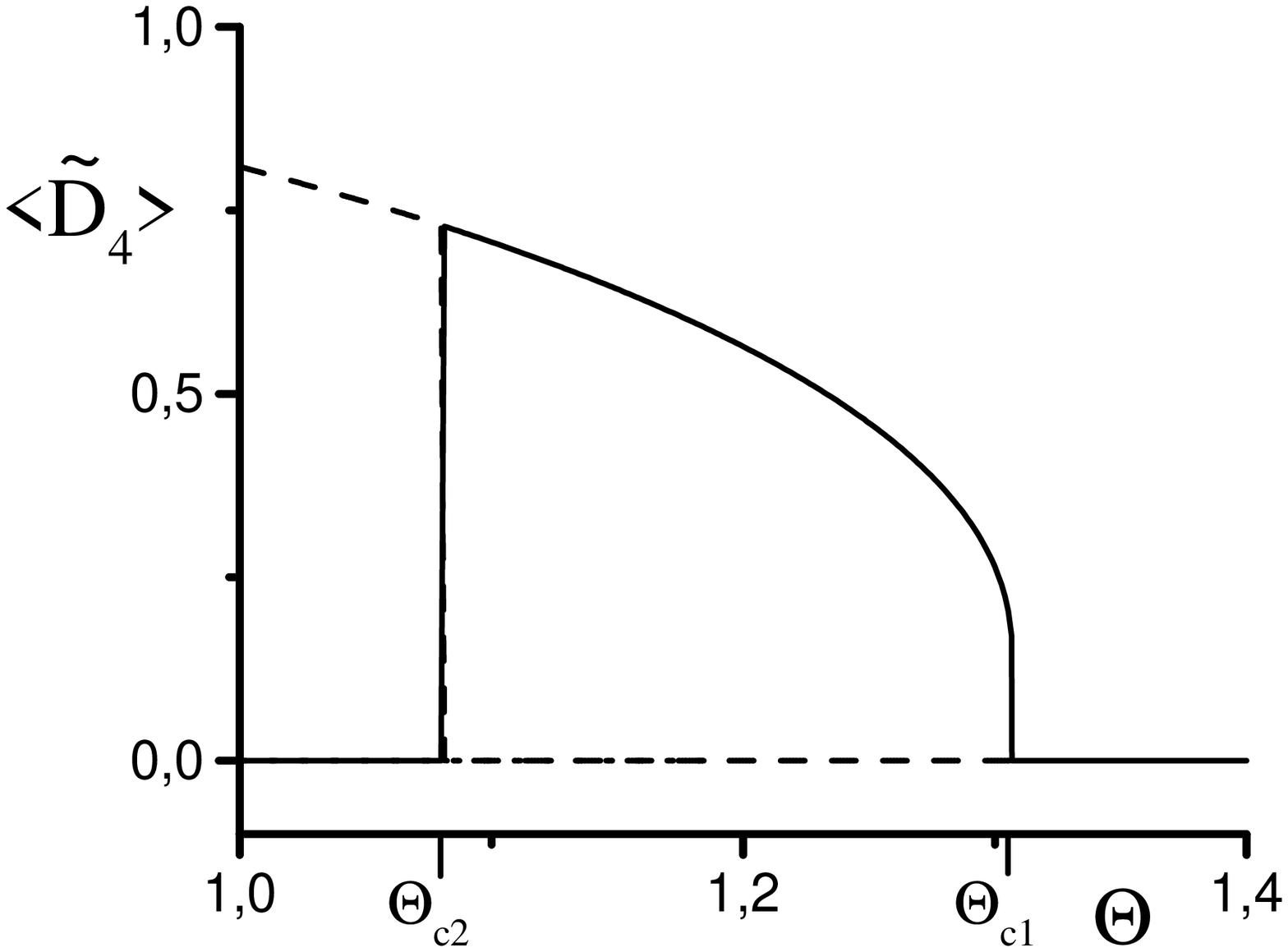}}
\put(5,0){\includegraphics[width=0.5\textwidth]{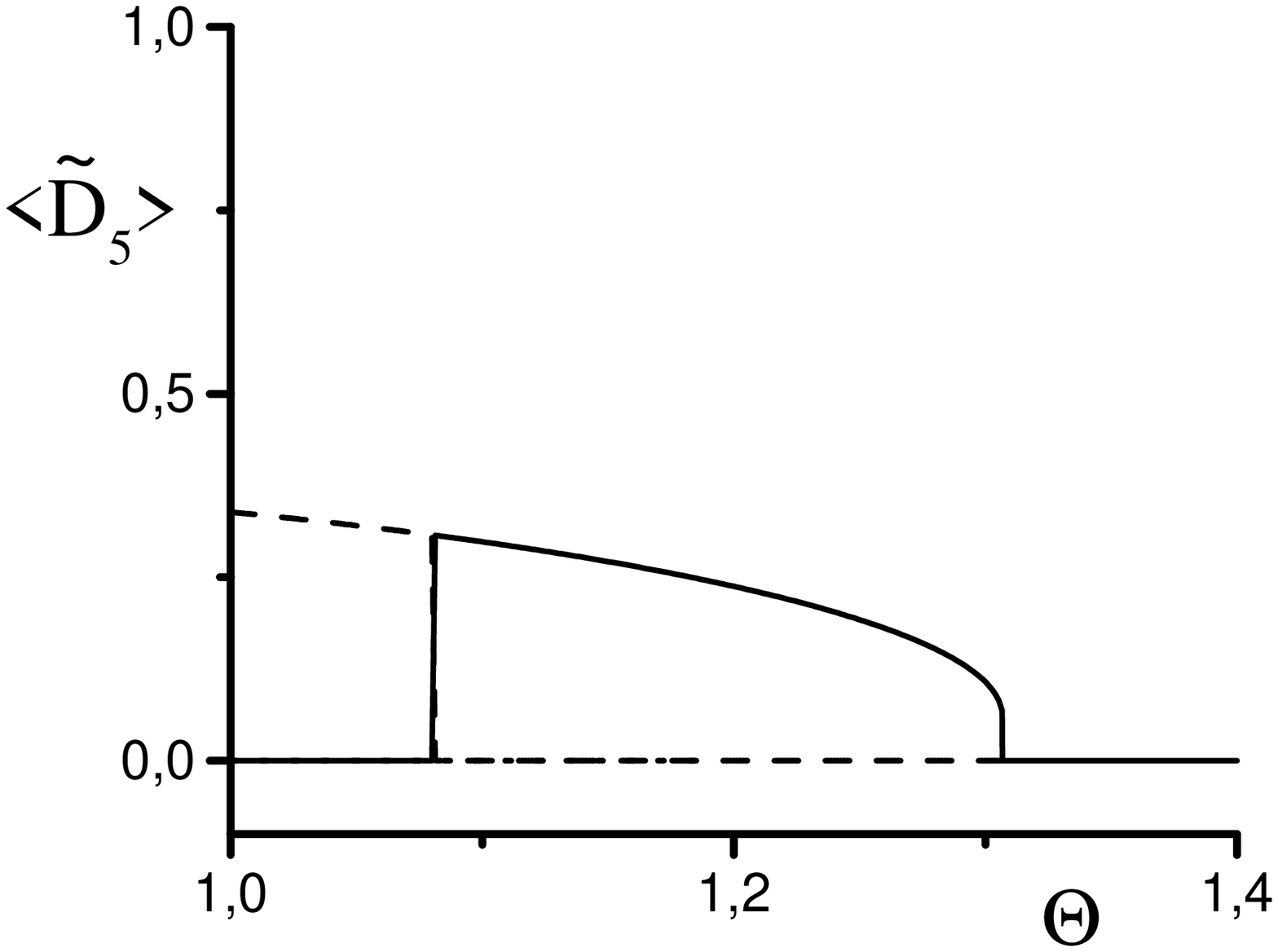}}
\end{picture}
\end{center}
\vspace{-2ex}

\caption{Dependences of $\langle\tilde{D}\rangle$ on temperature at
specified below values of parameters:  $b_1/a_1$=1.55, $d_1/a_1$=0.5,
$a_2/a_1$=0.4, $b_2/a_1$=2.6, $d_2/a_1$=1.6, $\Delta/a_1 d_x^2$=1.5855,
$\Omega_1$=$\Omega_1$=0, $d_y/d_x$ = 1.4; $\Theta$ is given in
$a_1 d_x^2$ units. The same values of parameters are held in
Figures~\ref{xt}--\ref{sy}. Solid lines correspond to the thermodynamically
stable solution, dashed and dotted -- to the unstable ones.}
\label{dtt}
\end{figure}

Results of numerical calculations of temperature dependences of symmetrized
dipole moments (\ref{eq12}), (\ref{eq13})  (obtained as solutions of
equation set (\ref{eq22})) having meaning of order parameters are depicted
in Figure~\ref{dtt}.  Here the case of the first order high-temperature
phase transition close to the second one is illustrated. Besides the
temperatures of real phase transitions $\Theta_{c1}$ and $\Theta_{c2}$, the
temperature of a possible phase transition $\Theta_{c}'$ (which might be
realized in the absence of the ferroelectric phase) is indicated.

\begin{figure}
\begin{center}
\setlength{\unitlength}{0.1\textwidth}
\begin{picture}(10,8)
\put(0,4){\includegraphics[width=0.5\textwidth]{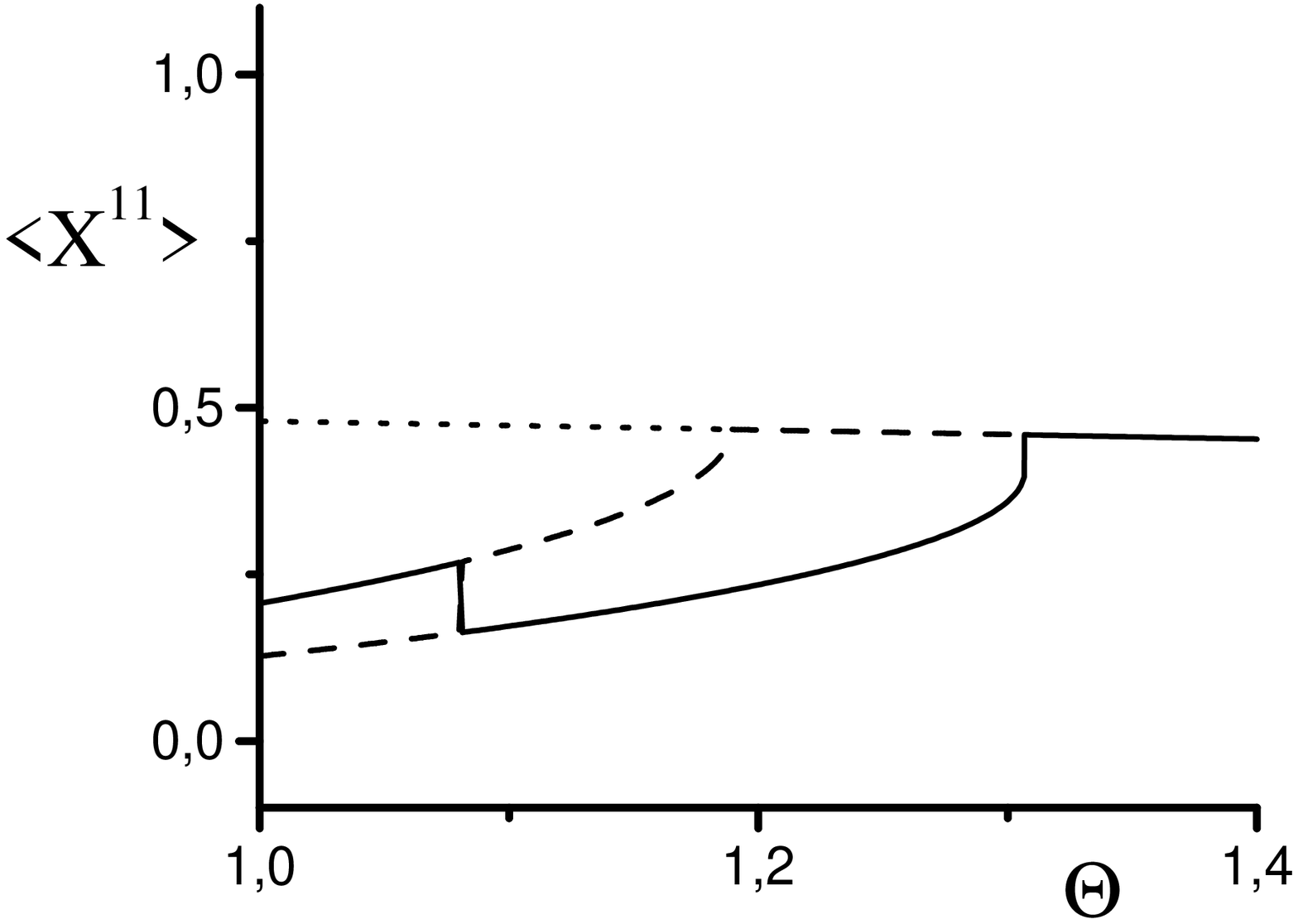}}
\put(5,4){\includegraphics[width=0.5\textwidth]{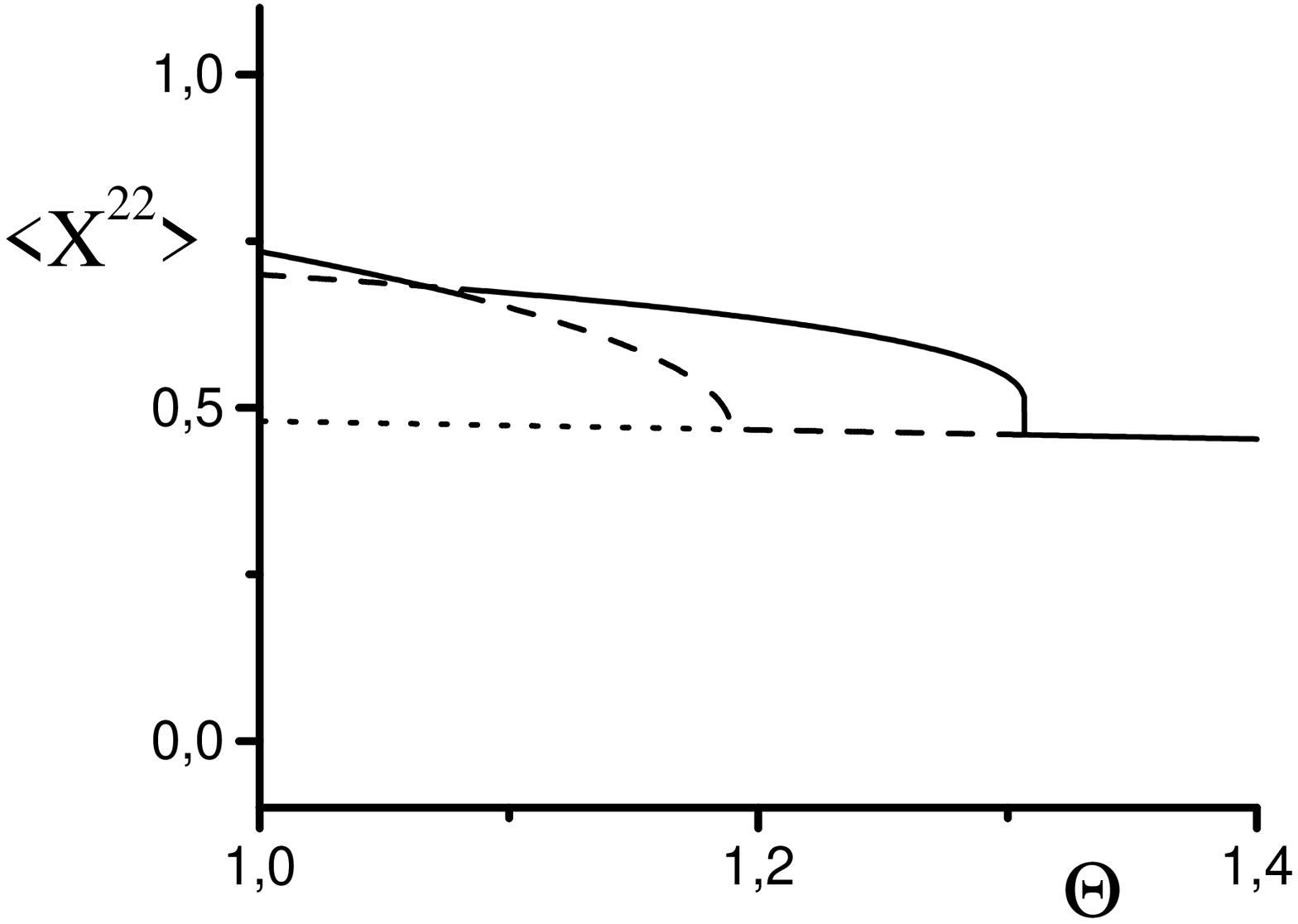}}
\put(0,0){\includegraphics[width=0.5\textwidth]{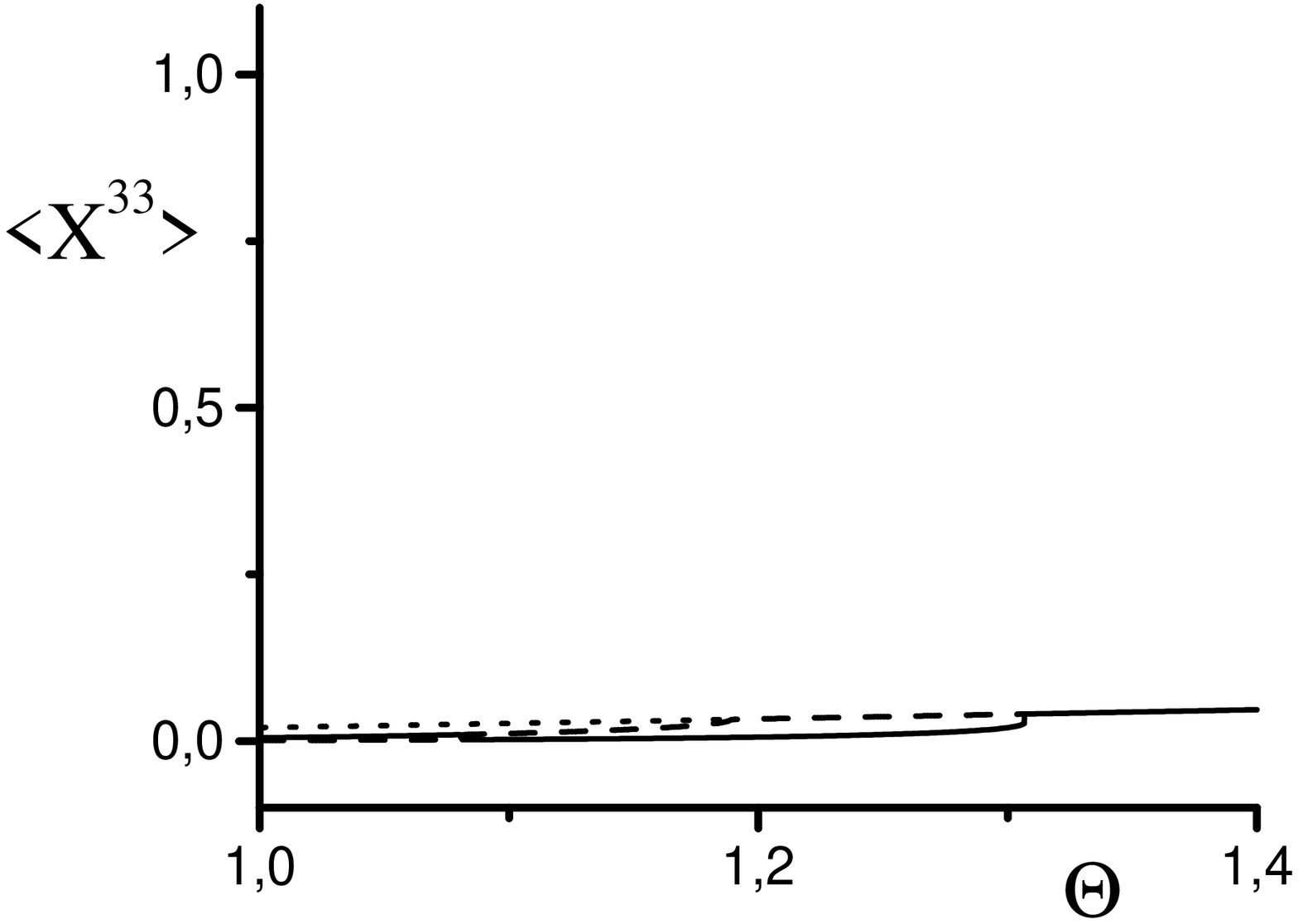}}
\put(5,0){\includegraphics[width=0.5\textwidth]{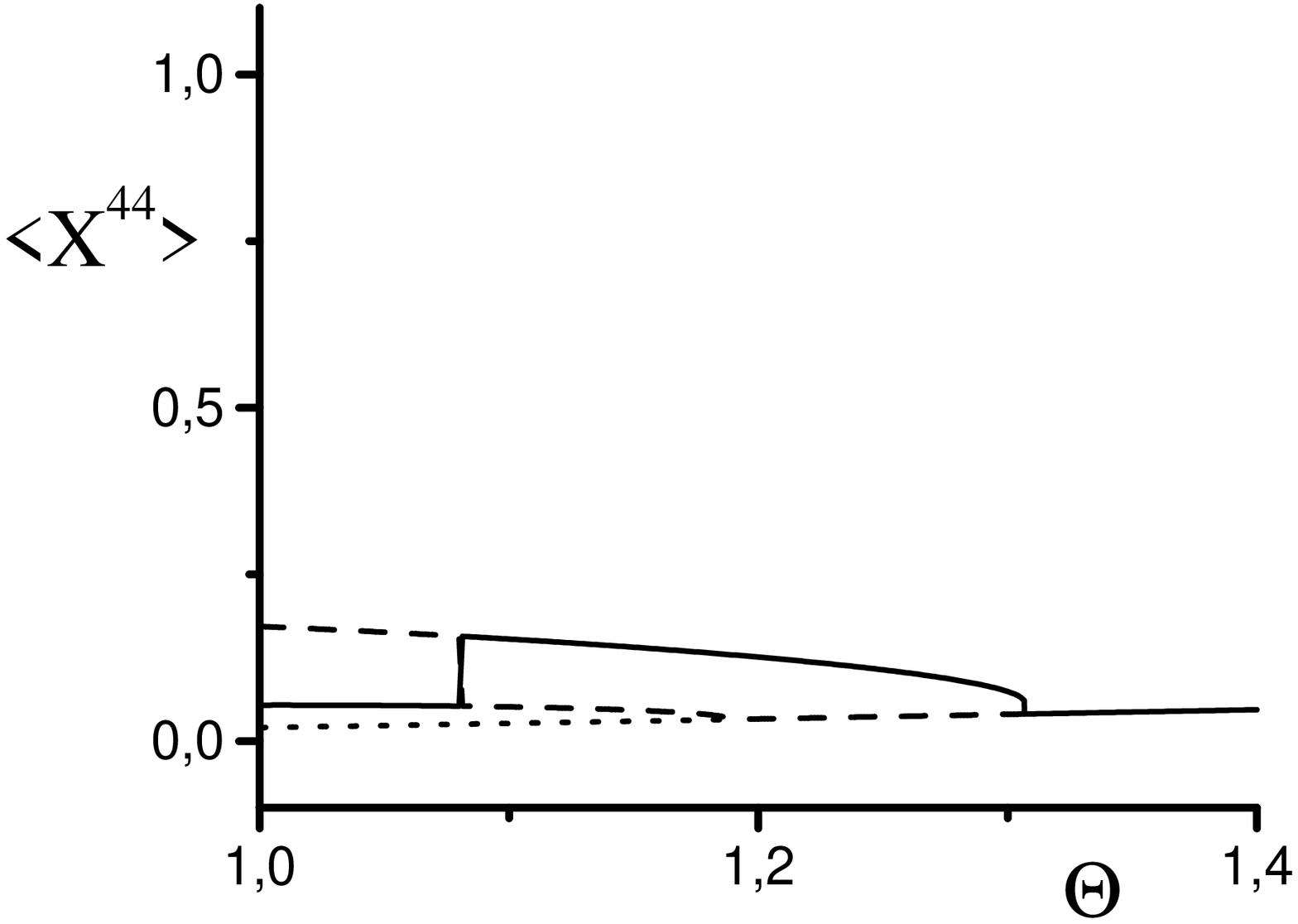}}
\end{picture}
\end{center}
\vspace{-2ex}

\caption{Dependences of occupancies of states on temperature. Here $k=1$.}
\label{xt}
\end{figure}

As one can see in Figure~\ref{dtt} and from the expressions (\ref{eq12}),
(\ref{eq13}) antiferroelectric orientation along the X axis is accompanied by
ferroelectric one along the Y axis and vice versa. Polarization along the X
axis is larger than along the Y one due to larger occupancies of positions 1
and 2 comparatively to positions 3 and 4 (Figure~\ref{xt}).

Behaviour of the spontaneous polarisation $P_x$ (which is
proportional to $\langle \tilde{D}_4 \rangle$ as stated in (\ref{scdef}))
strictly corresponds to experimental measurements (Dacko and Czapla, 1996;
Pykacz and Czapla, 1997). But despite of common assumption (see e.g.\
Kazimirov {\it et al.}, 1998) at lowering of temperature the occupancy of the
positions 3 and 4 do not tend to zero monotonically:
occupancies are asymmetrical and $\langle X^{44}_1 \rangle$ even increases
in the ferroelectric phase (with vice versa occupancies for $k=2$).

\begin{figure}
\begin{center}
\begin{tabular}{p{0.47\textwidth}p{0.47\textwidth}}
\strut
\hfill
\includegraphics[width=0.41\textwidth]{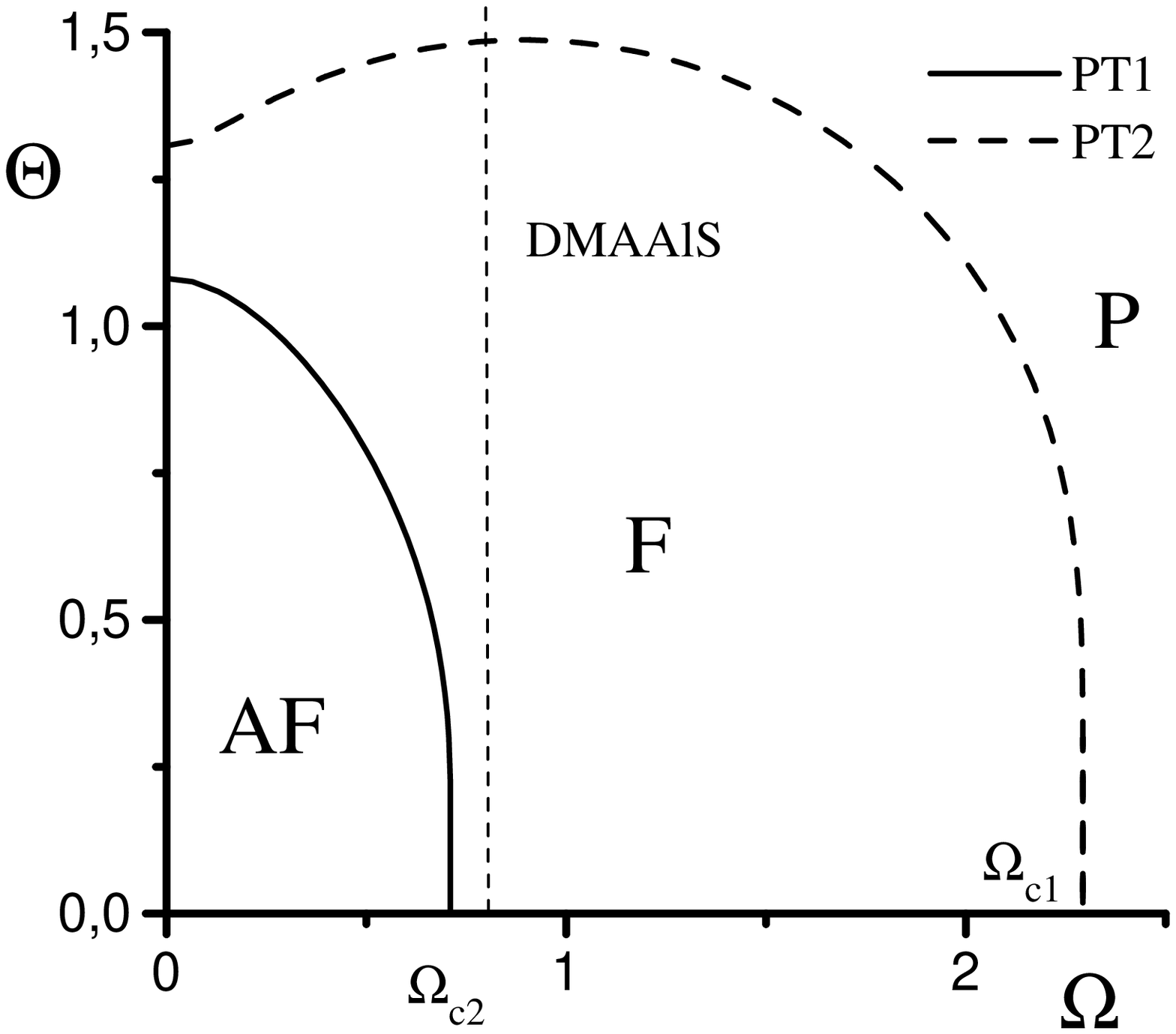}
\hfill
\strut
&
\strut
\hfill
\includegraphics[width=0.41\textwidth]{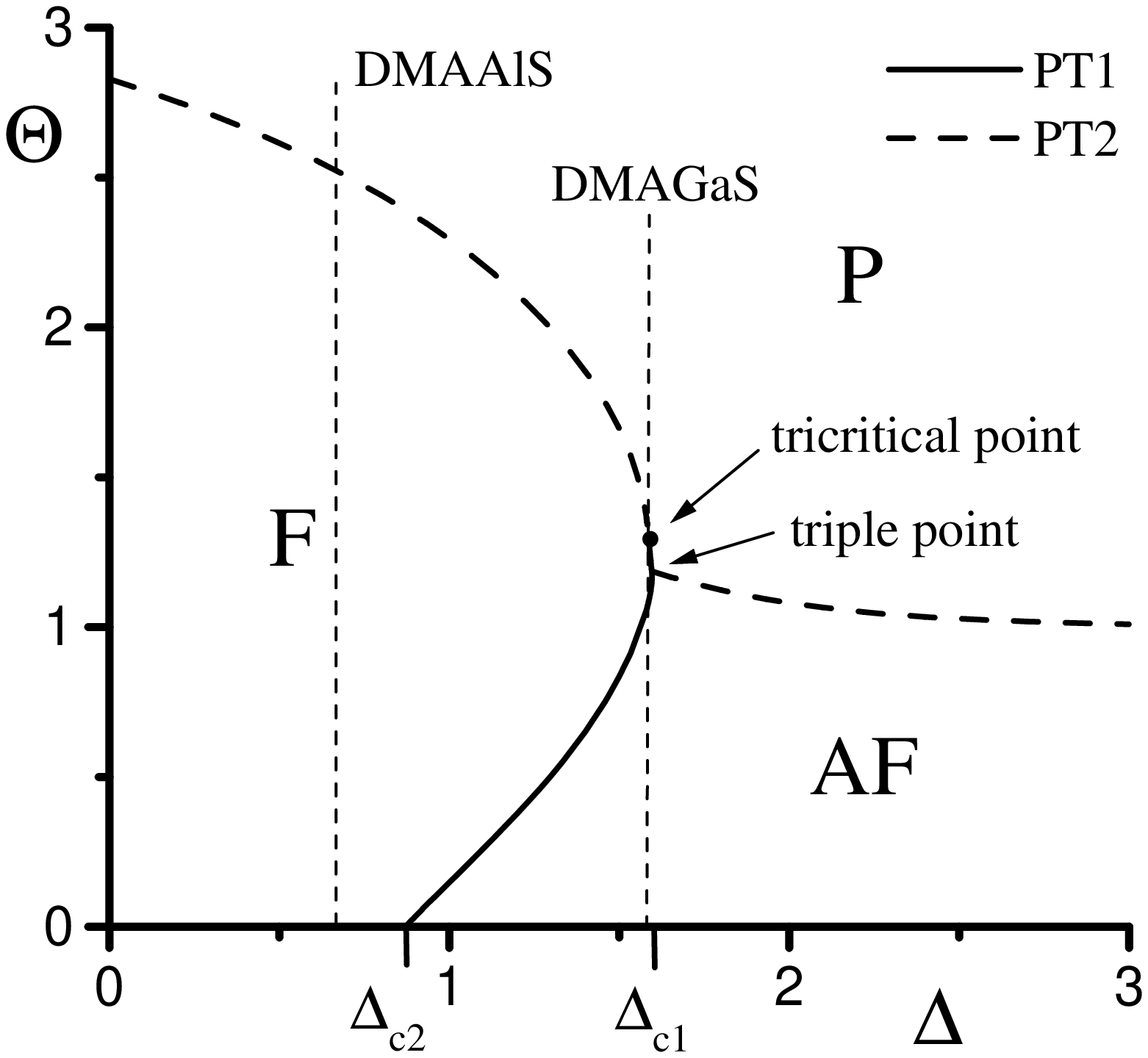}
\hfill
\strut
\\
\caption{%
Phase diagram $\Theta-\Omega$. $\Theta$ and $\Omega$ are given
$a_1 d_x^2$ units.
}
\label{ot}
&
\caption{%
Phase diagram $\Theta-\Delta$.
$\Theta$ and $\Delta$ are given $a_1 d_x^2$ units.
}
\label{dt}
\end{tabular}
\end{center}
\vspace{-5ex}
\end{figure}

The influence of the orientational hopping on the phase transition in the
system resemble one for Ising model (Figure~\ref{ot} -- for simplicity is
assumed $\Omega_{1}=\Omega_{2}=\Omega$). Namely the increase of $\Omega$
value leads as a whole to the decrease of temperatures of both phase
transitions and their sequential zeroing at corresponding values
$\Omega_{c2}$ and $\Omega_{c1}$.  Antiferroelectric phase vanishes at
relatively low values of $\Omega$ when the temperature of the other phase
transition changes insignificantly.  Due to prevailing suppression
of antiferroelectric ordering and the mentioned above competition between
anti- and ferroelectric interactions the considered diagram has a peculiar
feature:  the temperature of the ferroelectric-paraelectric phase transition
increases at small value of $\Omega$.

\begin{figure}
\includegraphics[width=0.45\textwidth]{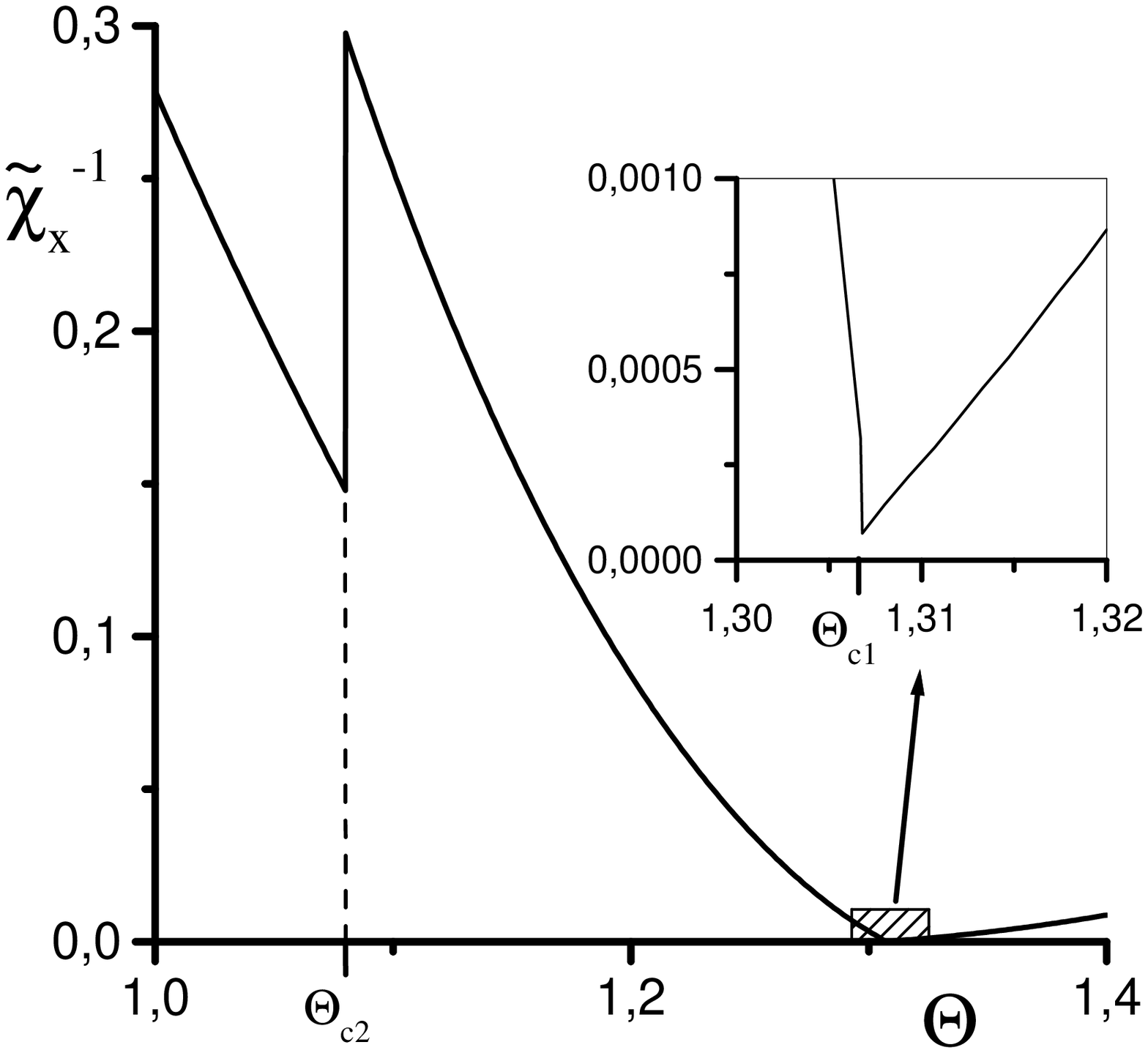}
\hfil
\includegraphics[width=0.45\textwidth]{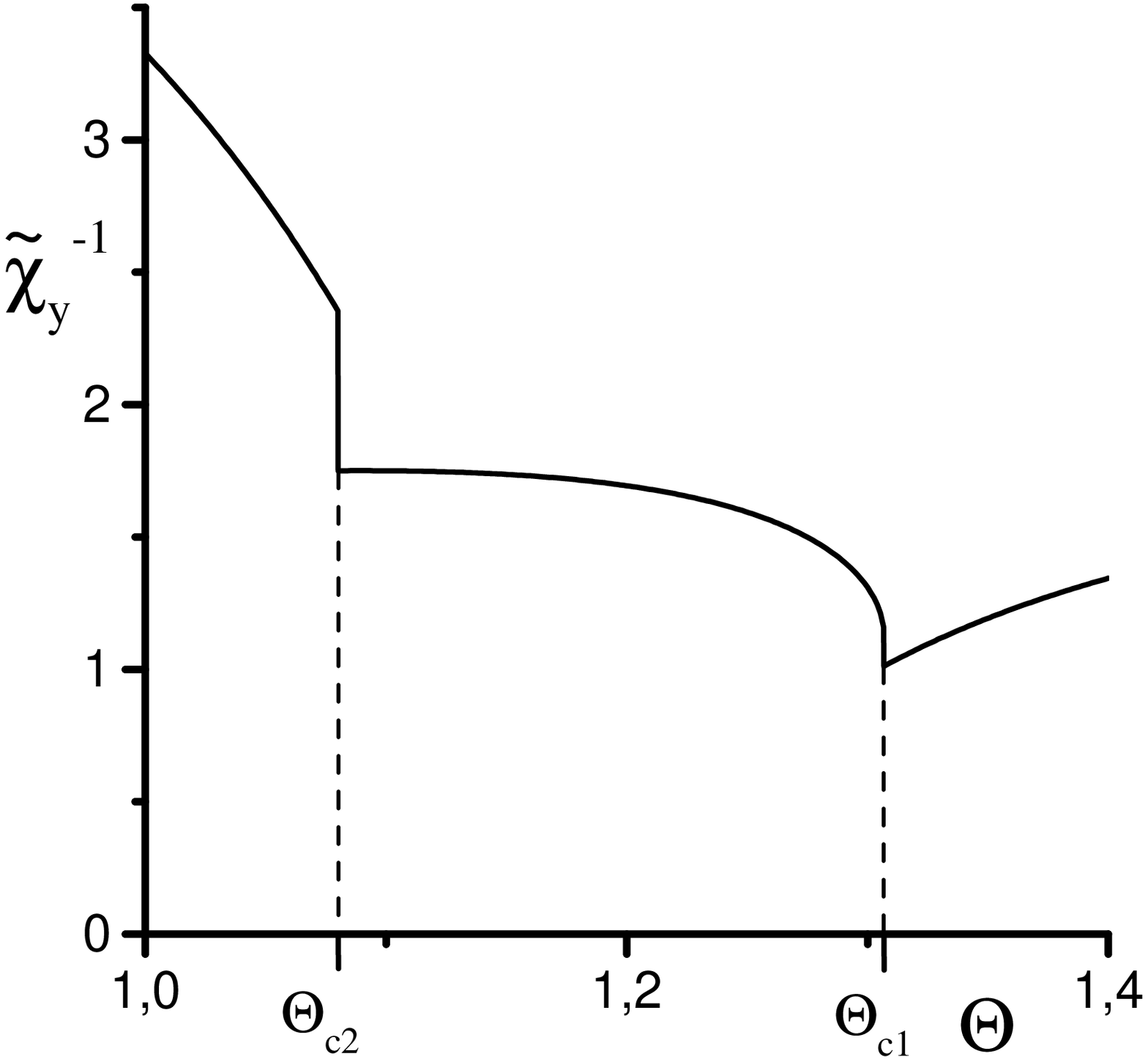}
\caption{Dependences of components of inverse static dielectrical
susceptibility on temperature. Here and in Figure~\ref{sy}
$\tilde{\chi}_{\alpha} = a_1 v_c \chi_{\alpha}$.}
\label{si}
\end{figure}

\begin{figure}
\centerline{\includegraphics[width=0.45\textwidth]{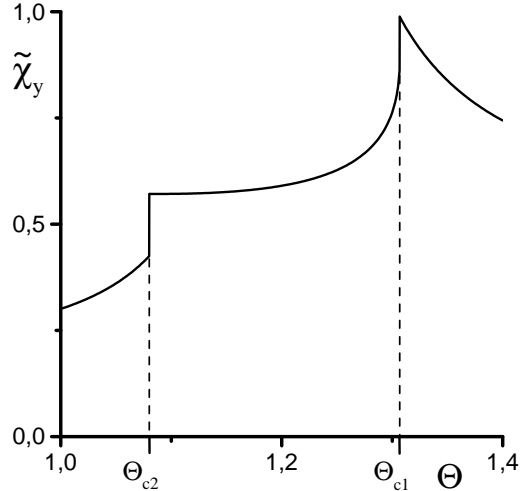}}
\caption{Dependence of the Y component of static dielectrical
susceptibility on temperature.}
\label{sy}
\end{figure}

According to the experimental data on pressure effects in DMAGaS (Yasuda,
Kaneda and Czapla, 1999) applying and increase of hydrostatic pressure leads to
the narrowing of the temperature region of existence of the ferroelectric
phase. At certain pressure the ferroelectric phase is completely suppressed.
Thermodynamical description in the framework of the Landau theory allows to
reproduce observed phase diagram by setting coefficients at the second order
terms in the Landau expansion of free energy to be pressure dependent
(Stasyuk, Velychko, Czapla and Czukwinski, 2000b). Microscopic
description of influence of hydrostatic pressure is rather a complicated
question. One can assume, that due to limited compressibility of the crystal
interaction constants change insignificantly and the main effect is due to
the increase of the difference of site energies $\Delta$. In this case the
obtained phase diagram remarkably good coincides with qualitative form of
the experimental one (Figure~\ref{dt}). All three
phases could exist at change of temperature in the region bounded by values
$\Delta_{c1}$ and $\Delta_{c2}$.
This confirms the above assumption of
prevailing influence of pressure on site energies. Nevertheless, it
seems that a more complete reproduction of the observed dependences (in
particular, the increase of temperature of phase transition in the
antiferroelectric phase at the pressure values above the triple point) can
be achieved when additional factors (such as change of interaction
constants or the parameter $\Omega$) will be taken into account.
For example, values of parameters of interaction are known to increase at
rising of pressure. The temperature $\Theta$ presented in
Figures~\ref{dtt}--\ref{sy} is made dimensionless by division on $a_1
d_x^2$. Hence if the dimensionless $\Theta_c$ remains constant at increase
of pressure, the real temperature of the phase transition grows up.

As one can mention there are both triple and tricritical points on the
diagram. There is a narrow region of $\Delta$ values near the triple point
where both phase transition are of the first order as in the case of DMAGaS
crystals.

Parameters in Figures~\ref{dtt} and \ref{xt} are chosen to fulfill
the mentioned above conditions and to give the best description of
experimental data for DMAGaS crystal. On the other hand, a typical behaviour
of the DMAAlS crystal could be obtained in two different way. One can
assume that value of the parameter $\Delta$ is smaller than $\Delta_{c2}$
for this case (Figure~\ref{dt}). Alternatively, value of the parameter $\Omega$
can be less than $\Omega_{c2}$ (Figure~\ref{ot}). In both cases it gives
direct transition of the second order from the paraelectric phase to the
ferroelectric one at lowering of temperature.

Dependences of inverse components of dielectric susceptibility are presented
in Figure~\ref{si}. $\chi_x(T)$ has a pronounced peak at the temperature of
phase transition between ferroelectric and paraelectric phases but does not
reach singularity.
Such a behaviour good coincides with the experimental measurements of
dielectric permittivity made by Tchukvinsky {\it et al.} (1998).
Due to the first order of the high-temperature phase
transition the slope of the $\chi_x(T)$ curve changes more than twice
in the phase transition point. $\chi_y(T)$ has a small peak in this point
(Figure~\ref{sy}).
There is a lack of experimantal data considering dielectric measurements
along the OY axis for the DMAGaS crystal but those for the DMAAlS
(Kapustianik {\it et al.}, 1994) close resemble the
theoretical picture.
Both components have jumps at temperature of the first order
phase transition from antiferroelectric to ferroelectric phase.

Basing on the experimental data some approximate evaluation of values of
parameters has been made. A value of polarization of saturation
$P_S=1.8\cdot10^{-2}$~C/m$^2$ gives $d_x=6.8\cdot10^{-30}$~C$\cdot$m
(what corresponds to 0.9e${}\times{}$0.5~\AA). Taking an approximate value
$a_1=27.4\cdot10^{20}$~cm$^{-3}$, we obtain $a_1 d_x^2=8.3\cdot10^{-3}$~eV
(normalizing factor for $\Theta$, $\Delta$ and $\Omega$) and
$a_1 v_c = 27.4\cdot10^{20}\;{\rm cm}^{-3}\times
           786\cdot10^{-24}\;{\rm cm}^{3} = 2.15$
(normalizing factor for the components of $\chi$). Hence we get
$\Delta=0.013$~eV (this value corresponds to the observed ratio of
occupancies of positions 1,2 and 3,4 at ambient temperature),
$\Omega_{c2} = 40$~cm$^{-1}$ and the maximum value of
$\chi_y = 13$ at the point of the high-temperature phase transition what
good coincides with the experiment for DMAAlS.

\section{Conclusions}

A simple dipole approximation for the four-state model proposed for
description of thermodynamics of the DMAGaS-DMAAlS family of crystals is
developed. The model exhibits complicated thermodynamical behaviour
depending on values of system parameters. At certain values of parameters
the model gives the same sequences of phase transitions as found
experimentally in DMAGaS and DMAAlS.

Numerical exploration of the model shows that increase of the
orientational hopping
between sites leads to the suppression of ordered phases in the Ising model
like way.

An attempt to give theoretical explanation of the experiment on the
influence of hydrostatic pressure on phase diagram of DMAGaS crystal results
in good qualitative agreement. It is assumed that the main effect of
pressure is the change of the difference of site energies.

\section{Acknowledgements}

The authors would like to thank Prof.~Z.Czapla and
Dr.~R.Tchuk\-win\-s\-kyi for their interest in the work and useful
discussions

This work was supported in part by the Foundation for
Fundamental Investigations of Ukrainian Ministry in Affairs of Science and
Technology, project No.~2.4/171.

\section{Appendix A. Transformation of the matrix of di\-po\-le-dipole
interactions to irreducible representations of the high-temperature symmetry
group}

\begin{equation}
\hat{\psi} = \left(\begin{array}{cccccc}
a' & b' & c' & d' & e' & f' \\
b' & g' & h' & -e' & k' & l' \\
c' & h' & m' & f' & -l' & n' \\
d' & -e' & f' & a' & -b' & c' \\
e' & k' & -l' & -b' & g' & -h' \\
f' & l' & n' & c' & -h' & m' \\
\end{array} \right).
\end{equation}
Here axis indices are changed first (e.g. $\{k\alpha\} =
\{1x,1y,1z,2x,2y,2z\}$).

Matrix of unitary transformation is as follows:
\begin{equation}
\hat{U} = \frac{1}{\sqrt{2}} \left(\begin{array}{cccccc}
1 & 0 & 0 & -1 & 0 & 0 \\
0 & 1 & 0 & 0 & 1 & 0  \\
0 & 0 & 1 & 0 & 0 & -1 \\
1 & 0 & 0 & 1 & 0 & 0  \\
0 & 1 & 0 & 0 & 1 & 0  \\
0 & 0 & 1 & 0 & 0 & 1  \\
\end{array}\right).
\end{equation}

After the transformation the matrix of interactions
\begin{eqnarray}
\hat{\tilde{\psi}} &=&
\frac{1}{2}
\left(
\begin{array}{cccccc}
a'-d' & b'+e' & c'-f' & 0 & 0 & 0 \\
b'+e' & g'+k' & h'-l' & 0 & 0 & 0 \\
c'-f' & h'-l' & m'-n' & 0 & 0 & 0 \\
0 & 0 & 0 & a'+d' & b'-e' & c'+f' \\
0 & 0 & 0 & b'-e' & g'-k' & h'+l' \\
0 & 0 & 0 & c'+f' & h'+l' & n'+m' \\
\end{array}
\right)
\nonumber\\
&=&
\hphantom{\frac{1}{2}}
\left(
\begin{array}{cccccc}
a_{1} & b_{1} & c_{1} & 0 & 0 & 0 \\
b_{1} & d_{1} & e_{1} & 0 & 0 & 0 \\
c_{1} & e_{1} & f_{1} & 0 & 0 & 0 \\
0 & 0 & 0 & a_{2} & b_{2} & c_{2} \\
0 & 0 & 0 & b_{2} & d_{2} & e_{2} \\
0 & 0 & 0 & c_{2} & e_{2} & f_{2} \\
\end{array}
\right)
\end{eqnarray}
becomes block-diagonal, where the upper left block corresponds to the
$A_{u}$ representation and the bottom right one to the $B_{u}$.

\end{document}